# Rheology modulated non-equilibrium fluctuations in time-dependent diffusion processes

Debonil Maity,[1] Aditya Bandopadhyay[2] & Suman Chakraborty[1,2]

[1]Department of Mechanical Engineering, Indian Institute of Technology Kharagpur, Kharagpur, India-721302

[2]Advanced Technology Development Center, Indian Institute of Technology Kharagpur, Kharagpur, India-721302

## HIGHLIGHTS

- The auto-correlation function (static and dynamic structure factor) of concentration fluctuations is determined for viscoelastic fluids
- The transverse component of velocity fluctuations which influences Rayleigh spectrum is obtained for viscoelastic fluids
- Criteria for the appearance of peaks at non-zero frequency for the case of dynamic structure factor is obtained for viscoelastic fluids
- Contrary to the equilibrium scenario the non-equilibrium Rayleigh line is influenced by viscoelastic effects

## ABSTRACT

The effect of non-Newtonian rheology, manifested through a viscoelastic linearized Maxwell model, on the time-dependent non-equilibrium concentration fluctuations due to free diffusion as well as thermal diffusion of a species is analyzed theoretically. Contrary to the belief that non-equilibrium Rayleigh line is not influenced by viscoelastic effects, through rigorous calculations, we put forward the fact that viscoelastic effects do influence the non-equilibrium Rayleigh line, while the effects are absent for the equilibrium scenario. The non-equilibrium process is quantified through the concentration fluctuation auto-correlation function, also known as the structure factor. The analysis reveals that the effect of rheology is prominent for both the cases of free diffusion and thermal diffusion at long times, where the influence of rheology dictates not only the location of the peaks in concentration dynamic structure factors, but also the magnitudes; such peaks in dynamic structure factors are absent in the case of Newtonian fluid. At smaller times, for the case of free diffusion, presence of time-dependent peak(s) are observed, which are weakly dependent on the influence of rheology, a phenomenon which is absent in the

_______________________________________________________________

Email address for correspondence: suman@mech.iitkgp.ernet.in

case of thermal diffusion. Different regimes of the frequency dependent overall dynamic structure factor, depending on the interplay of the fluid relaxation time and momentum diffusivity, are evaluated. The static structure factor is not affected to a great extent for the case of free-diffusion and is unaffected for the case of thermal diffusion.

## I. INTRODUCTION

A system at a finite temperature exhibits spontaneous and continuous microscopic fluctuations in quantities such as density, pressure, temperature, velocity field etc. about a mean value [1]. The fluctuations are dissipated in the system through various processes like diffusion, viscous flow or thermal conduction, much like the way in which external disturbances are relaxed. In course of this relaxation, each mode or degree of freedom of the system tries to return to the original state which can be a state of thermal equilibrium or non-equilibrium (which is assumed to be locally in thermal equilibrium). In the principle of local equilibrium, it is assumed that the macroscopic level can be divided into smaller systems which are infinitesimally smaller as compared to the macroscopic level but still contain sufficiently large number of molecules, so that statistical averages can be performed locally [2]. Notably, the local thermodynamic variables and other associated local thermodynamic properties remain interrelated by the same relations as for a thermodynamic equilibrium state.

Fluctuations in fluids in thermodynamic equilibrium have been studied to a great extent [2]. It is a very well-established fact that the correlation functions associated with the thermal fluctuations of fluid in equilibrium are spatially short ranged [2,3] (except for states near a critical point), unlike the case for fluids which are not in equilibrium. It is also well known that this long-ranged nature is caused by a coupling between hydrodynamic modes through the externally imposed non-equilibrium fluxes or gradients [3,4]. Thermal fluctuations are characterized by space-time functions that account for the correlation between the value of a quantity at some position and time with its value at a different position and time; these can be typically represented as a space-time correlation function $C_{AB}\left(\mathbf{x}_1, t_1; \mathbf{x}_2, t_2\right) = V\left\langle \delta A\left(\mathbf{x}_1, t_1\right) \delta B\left(\mathbf{x}_2, t_2\right)\right\rangle$ [1,2] where $A$ and $B$ denote the fluctuations in the dynamical variables under consideration, the subscripts denote two different states 1 and 2 respectively and the parameter $V$ denotes a multiplication prefactor. When $A$ and $B$ are the same variables, we obtain the autocorrelation function. When one considers the correlation function of density, one obtains the structure factor/density correlation function. The equal-time or static structure factor is defined as $S(k,0) = < \delta c_{k,0} \delta c_{k,0}^* >$ whereas the dynamic structure factor is defined as $S(k,\omega) = < \delta c_{k,\omega} \delta c_{k,\omega}^* >$ [2,5], where asterisk denotes the complex conjugate (via Fourier transform of the quantities). The intensity of the fluctuations is determined by equal-time correlation functions, whereas time-dependent correlation functions are used for determining the dynamical properties of fluctuations [2]. Several researchers have employed light scattering measurements as means to directly observe the evolution of the dynamic structure factor [1,2,6ó 8]. In these experiments, the wavenumber of observation is that associated with momentum transfer between the light and the sample during scattering which is approximately given by

$$\lambda = \frac{\lambda_0}{2}\sin\left(\theta/2\right) \quad [1],$$

where $\lambda$ is the scattered wavelength, $\lambda_0$ is the incoming wavelength of light and $\theta$ is the scattering angle (angle formed between the incoming and outgoing beams).

Recent developments in laser technology have led to the usage of He-Ne laser (6.328 pm) to measure the density correlation function for a wide variety of mixtures [1]. Steady state non-equilibrium fluctuations for a non-critical binary mixture in the presence of temperature gradient have been studied [7], where large enhancement to the so called Rayleigh line from concentration fluctuations, depending upon the square of the concentration gradient, has been predicted. Measurement of Soret and mass diffusion coefficients have been studied [9,10] by investigating non-equilibrium concentration fluctuations. Non-equilibrium fluctuations under the influence of gravity also have been studied in [11]. Fluctuations for isothermal free diffusion cases (the so called the non-Soret flux) have been studied by some researchers [5]. In such a process, the concentration evolves in space and time on a time scale much slower than the time scale of the fluctuations and hence, these processes can be considered to be quasi-stationary for fluctuations [5], while performing the Fourier transform for obtaining the structure factor. Thus, Vailati and Giglio [5] have obtained the static structure factor at any time $t$ by assuming a local equilibrium. For time dependent non-equilibrium diffusive processes, it is observed that the static structure factor shows a dramatic increase as compared to the value at equilibrium. However, for the case of non-equilibrium fluctuations at the steady state, the static structure factor depicts a $k^{-4}$ divergence at large wave numbers ($k$); these phenomena were also observed in experimental findings in the non-equilibrium fluctuations of binary mixture of aniline and cyclohexane [5]. Some experiments and theoretical works have showed the presence of unexpectedly large fluctuations in free diffusion processes [12ó14], which can be attributed to the coupling between the concentration and velocity fluctuations in non-equilibrium state [13].

While it is typical to consider an unbounded fluid domain for calculations, the importance of bounding length scales has been discussed in many works which deal with the finite sized effects for fluids or fluid mixtures bounded by boundaries [15ó19]. The asymptotic behaviour for certain range of dimensionless wave-numbers truly highlights the importance of accounting for the boundedness of the problem, which is manifested in the form of an intermediate length scale that affects the diffusion process. Through similar argument, in the case of non-Newtonian fluids, it is imperative to include the intermediate relaxation time scale of the fluid. This time scale is larger than the molecular relaxation time scale. Regardless, we expect that the relaxation time will play an important role towards dictating the structure factor at some time scale or the other. Some studies for equilibrium or non-equilibrium stationary states of complex łuids have been carried out [20,21]. Small-angle Rayleigh light-scattering measurements in polymer solutions under external temperature gradients have been performed and it has been reported that the concentration fluctuations are enhanced and are proportional to $\left(\nabla T\right)^2 / k^4$, where k in the wavenumber and $\nabla T$ is the external temperature gradient [22].

By theoretically considering the solution of Maxwell fluids, de Haro et al. have demonstrated [20] that although the Mountain peak [24626] is undisturbed by the viscoelastic nature of the fluid, the Brillouin peak (velocity correlations) becomes narrower and the asymmetry between the two Brillouin peaks increases [23]. On a separate note, the dynamic structure factor for a suspension of particles in a viscoelastic medium depicts alterations in the magnitude owing to the time-memory effects due to the relaxation time of the Maxwell fluid [27]. The presence of multiple relaxation time scales also leads to an enhancement in the concentration fluctuations; a notion which is associated with the so called resonance of the observation frequency and the fluid relaxation time. The memory effects of viscoelastic fluids may also be considered by taking fractional derivatives wherein the presence of the fractional derivative takes into account the memory of the past fluctuations and contributes to the present fluctuations via weighted residuals [27].

A central focus of the present work is to address the role of rheology towards the non-equilibrium fluctuations in the case of steady and unsteady concentration fluctuations, and that contrary to equilibrium situation, the non-equilibrium system contains the viscoelastic effects. Towards this, we analyze the system in which the rheology is defined by the linear Maxwell model, which provides us with analytical tractability. The central role of rheology is apparent at the Rayleigh spectrum. Firstly, we investigate the dynamic structure factor. In there we see that the interplay of the fluid relaxation time and momentum diffusivity time scale leads to dramatic consequences in the evaluation of the structure factor. We show that in both the cases of viscoelasticity and Newtonian curve (for the low and high frequency regime), the dynamic structure factor falls off as $\omega^{-2}$, whereas in the intermediate regimes, the viscoelastic curve decays faster than the Newtonian curve. The onset of the $\omega^{-2}$ regime for viscoelastic fluid is shifted towards lower frequencies; the magnitude of the structure factor is also higher. In the present work, we discuss the evaluation of structure factor for the case where $\nabla p = \rho \vec{g}$, which represents the variation of hydrostatic pressure due to gravity, where $p$ denotes the macroscopic pressure, $\rho$ denotes the macroscopic density, and $\vec{g}$ represents the acceleration due to gravity. We consider both isothermal free diffusion and thermal diffusion of the species in this work. At short time instances, for the case of free diffusion we observe the presence of time-dependent phenomena, which are primarily due to the coupling between gravity and concentration gradient. For the case of static structure factor, we observe that $k^{-4}$ behavior is retained for the case of thermal diffusion, while the curve decays slowly with respect to $k$ for the case of free diffusion and at short times. At longer times, when the concentration gradient saturates, $k^{-4}$ behavior is again retained for all cases (including non-Newtonian fluid mixtures). Furthermore, we come up with an analytical expression of the criteria for the appearance of peaks for dynamic structure factors for both the cases of absence of gravity and presence of gravity for large time instances.

## II. MATHEMATICAL FORMULATION

Non-equilibrium fluctuations in diffusion processes are obtained by using the fluctuating hydrodynamics equations, along with phenomenological equations describing the variation of the macroscopic state of a Maxwell fluid mixture [23]. We have neglected temperature fluctuations in this case, so that the relevant hydrodynamic variables are the density $\rho$, the concentration $c$, and the velocity $\vec{u}$ of the fluid. Further we assume that the mixture is at rest. Considering these, the transport equations, under the action of gravity, become

$$\frac{\partial c}{\partial t} + \vec{u}.\nabla c = -\frac{1}{\rho}\nabla.\vec{j} \tag{1}$$

$$\vec{j} = -\rho D\left(\nabla c + \frac{k_T}{T}\nabla T + \frac{k_p}{p}\nabla p\right) \tag{2}$$

$$\rho\frac{d\vec{u}}{dt} = -\nabla p - \nabla.\vec{\vec{\tau}} + \rho\vec{g} \tag{3}$$

$$-\tau_r\frac{d\vec{\vec{\tau}}}{dt} = \vec{\vec{\tau}} + \eta\left[(\nabla\vec{u}) + (\nabla\vec{u})^T\right] \tag{4}$$

where $\vec{j}$ is the mass flux, $p$ is the hydrostatic pressure, $\vec{\vec{\tau}}$ is the deviatoric stress tensor for the Maxwell fluid, $k_T$ is the Thermal diffusion coefficient, $k_p$ is the baro-diffusion coefficient, $D$ is the diffusion coefficient, $\eta, \eta_\nu$ represent the dynamic viscosity and bulk viscosity respectively, $d/dt$ represents the total derivative defined as $\frac{d}{dt} \equiv \frac{\partial}{\partial t} + \vec{u}.\vec{\nabla}$, $t$ is the time, $\tau_r$ is the Maxwell relaxation time and $\vec{\vec{1}}$ is the unit tensor. We use the linearized Maxwell model, for which the constitutive behavior is given by Eq. (4). It is to be noted that the linearized Maxwell model as described is a much simplified form of the more general framework which is described by the Phanó–Thienó–Tanner (PTT) model [31,32]. The approximation of a linearized Model does not always hold true at very large shear/strain rates, however, the PTT model is not amenable to analytical treatment for the physical problem at hand. Therefore, by employing the linearized Maxwell model, we are able to retain the analytical tractability of the problem without sacrificing most of the underlying physics. We may also note that the linearized Maxwell model has been employed in some of the earlier works pertaining to non-equilibrium fluctuations [23,27] in order to make qualitative assessments of the underlying phenomena.

Proceeding to decompose the variables as a mean and fluctuating part, the variables can be written as follows:

$$\rho(\vec{r},t) = <\rho(\vec{r},t)> + \delta\rho(\vec{r},t) \tag{5}$$

$$\vec{j} = <\vec{j}(\vec{r},t)> + \delta\vec{j}(\vec{r},t) \qquad (6)$$

$$\vec{u}(\vec{r},t) = <\vec{u}(\vec{r},t)> + \delta\vec{u}(\vec{r},t) = \delta\vec{u}(\vec{r},t) \qquad (7)$$

$$c(\vec{r},t) = <c(\vec{r},t)> + \delta c(\vec{r},t) \qquad (8)$$

$$\vec{\vec{\tau}}(\vec{r},t) = \delta\vec{\vec{\tau}}(\vec{r},t) \qquad (9)$$

In general, we may denote the density fluctuation as being comprised of a species fluctuation component, a pressure fluctuation component and the temperature fluctuation component. Therefore in Eq. (5) we can write $\delta\rho(\vec{r},t) = \rho[\beta\delta c + \beta_p\delta p - \alpha_T\delta T]$, where, $\beta = \rho^{-1}(\partial\rho/\partial c)_{p,T}$, $\beta_p = \rho^{-1}(\partial\rho/\partial p)_{T,c}$ and $\alpha_T = -\rho^{-1}(\partial\rho/\partial T)_{p,c}$. However we will assume that only concentration fluctuations are responsible for density fluctuations [5] and that the equilibrium condition is indicated by a fluid at rest, i.e. $<\vec{u}(\vec{r},t)> = 0$. For further analysis, the brackets around the variables would be dropped. Replacing Eqs. (5)-(9), in Eqs. (1)-(4) and assuming $|\delta Q| << |Q|$, where $Q$ is any generic quantity such as density, concentration and so on, we obtain the following equations represented in terms of the flux $\vec{j}$ :

$$\frac{\partial(c+\delta c)}{\partial t} + \delta\vec{u}.\nabla c = -\frac{1}{\rho}\nabla.\vec{j} - \frac{1}{\rho}\nabla.\delta\vec{j} + \frac{\beta\delta c}{\rho}\nabla.\vec{j} + \nabla.\vec{F} \qquad (10)$$

$$\frac{\partial\delta\vec{u}}{\partial t} = \beta\delta c\vec{g} - \frac{\nabla.\delta\vec{\vec{\tau}}}{\rho} + \frac{\nabla.\vec{\vec{\sigma}}}{\rho} \qquad (11)$$

$$-\tau_r\frac{\partial\delta\vec{\vec{\tau}}}{\partial t} = \delta\vec{\vec{\tau}} + \eta\left[(\nabla\delta v) + (\nabla\delta v)^T\right] \qquad (12)$$

where the random forces $\vec{F}$ and $\vec{\vec{\sigma}}$ had been added to describe the spontaneous onset of concentration and velocity fluctuations, respectively. The macroscopic variables are assumed to obey the Eqs.(1) to (4) with the relevant equilibrium conditions (such as no-flow). Under these assumptions, we obtain:

$$\frac{\partial c}{\partial t} + \nabla.\vec{j} = 0 \qquad (13)$$

$$\nabla p = \rho\vec{g} \qquad (14)$$

For simplicity, we will assume that gradients of the thermodynamic variables are small, so that we can neglect the spatial dependence of the thermo-physical properties of the mixture and macroscopic convection is absent.

Using Eq. (2), the fluctuating part of the mass flux is obtained as,

$$\nabla . \delta \vec{j} = -\rho D \nabla^2 \delta c + \beta \delta c \nabla . \vec{j} + \frac{\nabla \delta \rho . \vec{j}}{\rho} \tag{15}$$

In our further calculations, we will assume that macroscopic concentration and temperature gradients are parallel to the $z$-axis, and the wave vector $\vec{k}$ is perpendicular to these gradients [5]. As a result, the spatio-temporal Fourier transform would have the form,

$$\delta q_{k,\omega} = \int dt \int d\vec{r}\, q(\vec{r},t) \exp[i(\vec{k}.\vec{r} - \omega t)] \tag{16}$$

where $q$ may be $c$ or $\vec{u}$; the microscopic quantity (fluctuation) is transformed into the $(k,\omega)$ space as denoted by the subscript. During this calculation of the Fourier transform, the last two terms of Eq. (15) cancel out. The macroscopic variables are not affected by temporal Fourier transform as it can be assumed that frequencies associated with them are much smaller than those associated with the fluctuations [30]. Applying Fourier Transform to Eqs.(10), (11), and (12), supplemented by Eqs.(13), (14) and (15), and considering only the case of Rayleigh spectrum (which are solely influenced by transverse velocity fluctuations), we get (It is to be noted that we are keeping the transforming in $x$ and $y$ and hence, $\nabla c(z,t)$ is untransformed, which essentially makes the fluctuating variables dependent on $z$ and t [5]):

$$\delta c_{k,\omega}(i\omega + Dk^2) = -\delta \vec{u}_{k,\omega}.\nabla c - i\vec{k}.\vec{F}_{k,\omega} \tag{17}$$

$$\delta \vec{u}_{k,\omega}(i\omega) = \beta \vec{g} \delta c_{k,\omega} + \frac{i\vec{k}.\vec{\tau}_{k,\omega}}{\rho} - \frac{i\vec{k}.\vec{\sigma}_{k,\omega}}{\rho} \tag{18}$$

$$i\vec{k}.\vec{\tau}_{k,\omega} = -\frac{(1-i\omega\tau_r)}{(1+\omega^2\tau_r^2)}\Big[\eta k^2 \delta \vec{u}_{k,\omega}\Big] \tag{19}$$

Combining Eqs.(17-19), we obtain,

$$(\delta \vec{u}_{k,\omega}.\nabla c) = \frac{\beta(\vec{g}.\nabla c)\delta c_{k,\omega} - \dfrac{i(\vec{k}.\vec{\sigma}_{k,\omega}).\nabla c}{\rho}}{(i\omega + \eta X k^2)} \tag{20}$$

where, $X = \dfrac{(1-i\omega\tau_r)}{\rho(1+\omega^2\tau_r^2)}$

Combining Eqs.(18) and (20), we obtain,

$$\delta c_{k,\omega}(i\omega + Dk^2) = \frac{-\beta \delta c_{k,\omega}(\vec{g}.\nabla c)}{(i\omega + \eta X k^2)} + \frac{i\vec{k}.\vec{\sigma}_{k,\omega}.(\nabla c)}{\rho(i\omega + \eta X k^2)} - i\vec{k}.\vec{F}_{k,\omega} \tag{21}$$

On simplifying, Eq. (21) becomes,

$$\delta c_{k,\omega} = \frac{+\dfrac{i\vec{k}.\vec{\sigma}_{k,\omega}.\nabla c}{\rho} - i\vec{k}.\vec{F}_{k,\omega}(i\omega + \eta X k^2)}{[(i\omega + D k^2)(i\omega + \eta X k^2) + \beta \vec{g}.\nabla c]} \quad (22)$$

The presence of only the transverse velocity fluctuation enforces that $\eta_v$ (bulk viscosity) is absent in the expression of the structure factor. The next step is to calculate the correlation functions for the fluctuations. For that we will assume that the correlations of the random forces retain their equilibrium values [5,7,23]. Also, for our system the divergence of velocity is zero. Accordingly,

$$\left\langle F^i_{\vec{k},\omega} F^{*j}_{\vec{k}',\omega'} \right\rangle = \frac{2k_B T}{(2\pi)^4 \rho} D \left( \frac{\partial c}{\partial \mu} \right)_{p,T} \delta_{ij} \delta\left(\vec{k} - \vec{k}'\right) \delta\left(\vec{\omega} - \vec{\omega}'\right) \quad (23)$$

$$\left\langle \sigma^{ij}_{\vec{k},\omega} \sigma^{*lm}_{\vec{k}',\omega'} \right\rangle = \frac{2k_B T}{(2\pi)^4 \left(1 + \omega^2 \tau_r^2\right)} \left[ \eta \left( \delta_{il}\delta_{jm} + \delta_{im}\delta_{jl} \right) \right] \delta\left(\vec{k} - \vec{k}'\right) \delta\left(\vec{\omega} - \vec{\omega}'\right) \quad (24)$$

$$\left\langle F^i_{\vec{k},\omega} \sigma^{*lm}_{\vec{k}',\omega'} \right\rangle = 0 \quad (25)$$

Eq. (25) follows from Curie s principle which states that correlations among the components of the thermodynamic forces having different tensorial character is 0 [2]. (It is important to note that typically for equation [24], there have been attempts for the expression of dissipation [31] which is connected to entropy production and we use the expression which had already been assumed in the literature such as [23]). Using Eqs.(22-25), we may write the dynamic structure factor as:

$$< \delta c_{k,\omega} \delta c^*_{k,\omega} >_R = \frac{2k_B T}{(2\pi)^4} k^2 \left[ \frac{\dfrac{|(i\omega + \eta X k^2)|^2 D}{\rho} \left( \dfrac{\partial c}{\partial \mu} \right)_{p,T} + \dfrac{(\eta\,|\nabla c|^2)}{(1 + \omega^2 \tau_r^2)\rho^2}}{|(i\omega + D k^2)(i\omega + \eta X k^2) + \beta \vec{g}.\nabla c|^2} \right] \quad (26)$$

Eq. (26) represents the structure factor for non-equilibrium concentration fluctuations in a Maxwell fluid, $k$ is the wave number. In this respect, Rayleigh number ratio can be defined as: $-\dfrac{\rho \beta \vec{g}.\nabla c}{\eta D k^4}$. It is defined in such a way such that concentration gradient points along the direction of $\vec{g}$, since we want to avoid the situation which leads to the onset of convective instabilities [5]. For the case of thermal diffusion, we assume that the temperature gradient is applied by heating the layer from above, which avoids the onset of convective instabilities.

If the fluid is assumed to be Newtonian with $\tau_r = 0$, the non-dimensional number $X$ gets modified as $1/\rho$, and therefore Eq. (26) reduces to the limiting condition of the Rayleigh spectrum [5].

*Overall Dynamic Structure Factor* - We now investigate the nature of the overall dynamic structure factor. Experimental observations pertaining to the dynamic structure factor are done by shining a beam of light of a particular wavelength. The scattered light then arises due to fluctuations in the refractive index which in turn is related to the density fluctuations [2]. The density fluctuations can be expressed as a function of pressure, temperature and concentration fluctuations as $\delta\rho(\vec{r},t) = \rho[\beta\delta c + \beta_p\delta p - \alpha_T\delta T]$. A scattering setup is considered where the probe beam is aligned with the gravity [5]. The angular distribution of the scattered intensity changes layer by layer, therefore, the overall scattering distribution results from a summation of the scattered intensity distributions from the individual layers. The spectral density of scattered light for the Rayleigh spectrum is [5],

$$I(k,\omega) = \frac{Ik_i^4}{16\pi^2 R^2 \varepsilon_0^2}\left(\frac{\partial\varepsilon}{\partial c}\right)_{p,T}^2 S(k,\omega) \tag{27}$$

where $\varepsilon_0$ is the dielectric constant of the sample, I is the intensity of incident beam, $k_i$ is the magnitude of the incident wave vector and $S(k,\omega)$ is the overall dynamic structure factor, defined as,

$$S(k,\omega) = \left\langle \delta c_{k,\omega}\delta c_{k,\omega}^* \right\rangle \tag{28}$$

Here, $\delta c_{k,\omega}$ is the three-dimensional spatial Fourier transform of the concentration fluctuations. In our presentation we initially perform a 2 dimensional transform and later integrate the entire structure factor to take into account the fluctuations at all values of the untransformed co-ordinate.

Towards quantifying the structure factor found in Eqs.(26), we consider the following two cases: 1. Isothermal free diffusion and 2. Thermal diffusion.

*1. Isothermal Free Diffusion* - We consider here a case when two miscible liquids are initially separated by a distinct horizontal boundary. Hence the initial conditions are,

$$c(z,0) = \begin{matrix} c_1, 0 < z < h \\ c_2, h < z < a \end{matrix} \tag{29}$$

Where $c_1$ and $c_2$ are the concentration of fluids in the two respective layers, $h$ is the position of initial boundary between the two fluids and $a$ is the net thickness of the two layers of the fluids. The concentration profile can be solved using the above boundary conditions and by considering the free diffusion equation as had been pointed out in [5] and [33], so that the resultant profile becomes:

$$c(z,t) = \frac{c_1 h + c_2(h-a)}{a} + \frac{2}{\pi}(c_1 - c_2)\sum_{n=1}^{\infty}\frac{1}{n}\sin\left(\frac{n\pi h}{a}\right)\cos\left(\frac{n\pi z}{a}\right)\exp\left(-\frac{Dn^2\pi^2}{a^2}t\right) \qquad (30)$$

Just for the case of free diffusive regime (early stages of the free diffusion process), $\nabla c$ is given by, (please refer appendix figure A3)

$$|\nabla c(z,t)| = \frac{(c_1 - c_2)}{\sqrt{4\pi Dt}}\exp\left[-\frac{(z-h)^2}{4Dt}\right] \qquad (31)$$

While evaluating Eqs. (31) and (32), it has been assumed that the actual concentration gradient present in the mixture is much greater than the baro-diffusion term in Eq. (2).

2. *Thermal Diffusion*- According to [34], when a temperature gradient is applied to a fluid, a macroscopic mass flux is produced. This effect is known as thermal diffusion or Soret effect. The steady state concentration gradient in the absence of baro-diffusion would be,

$$\nabla c_{steady} = \nabla c_{soret} = -\frac{k_T}{T}\nabla T \qquad (32)$$

Furthermore, the boundary conditions are,

$$c = c_0, 0 \leq z \leq a, t = 0 \qquad (33)$$

$$\nabla c = \nabla c_{soret}, z = 0; a, t > 0 \qquad (34)$$

where $c_0$ is the initial concentration of the sample. We will apply the assumption that the mass flux is primarily dominated by thermal diffusion. We will stick to the assumption as had been made in [5], that the thermalization of the mixture is attained almost instantaneously with respect to the time needed to reach the steady concentration profile. Also, for the boundary condition (Eq. (35)), the concentration gradient must reach steady-state instantaneously where the mass flux is zero [5], in this case, the boundaries. The concentration profile can hence be solved as,

$$c(z,t) = c_0 + a|\nabla c_{steady}|\left[\frac{1}{2} - \frac{z}{a} - \frac{2}{\pi^2}\sum_{n=1}^{\infty}\frac{1}{n^2}\left(1-(-1)^n\right)\cos\left(\frac{n\pi z}{a}\right)\exp\left(-\frac{Dn^2\pi^2}{a^2}t\right)\right] \qquad (35)$$

$$|\nabla c(z,t)| = \left| a\nabla c_{steady} \left( -\frac{1}{a} + \frac{2}{\pi a} \sum_{n=1}^{\infty} \frac{1}{n}(1-(-1)^n)\sin\left(\frac{n\pi z}{a}\right)\exp\left(-\frac{Dn^2\pi^2}{a^2}t\right) \right) \right| \qquad (36)$$

## III. RESULTS AND DISCUSSIONS

In the following discussion, we have used the following values of the parameters [5,7,23]:

$\eta$ =0.000553 $Pa.s$, $\eta_v$ =0.00747 $Pa.s$, $D$=$10^{-10}$ $m^2/s$, $\beta$ =0.27, height of the vessel containing the liquid mixture, $a$=0.004 $m$, $\left(\dfrac{\partial c}{\partial \mu}\right)_{p,T}$ =$10^{-3}$ $s^2/m^2$, $k$ =85200 $m^{-1}$ and $2\times10^5$ $m^{-1}$, $\rho$ =860 $kg/m^3$.

Apart from these, for the two cases of free diffusion and thermal diffusion, we have employed the following parameters:

a) Free diffusion: For this case, we consider an arrangement where two horizontal layers of the binary mixture at the uniform concentrations, $c_1$=0.75 and $c_2$=0.25 are separated by a horizontal interface at the mid-height $a/2$ (=0.002 $m$).

b) Thermal Diffusion: The fluid mixture is assumed to be an initial concentration $c_0$=0.5, the temperature gradient $|\nabla T|$=16000 $K/m$, thermal diffusion ratio, $k_T$ =3.5, the temperature $T$ =315 $K$.

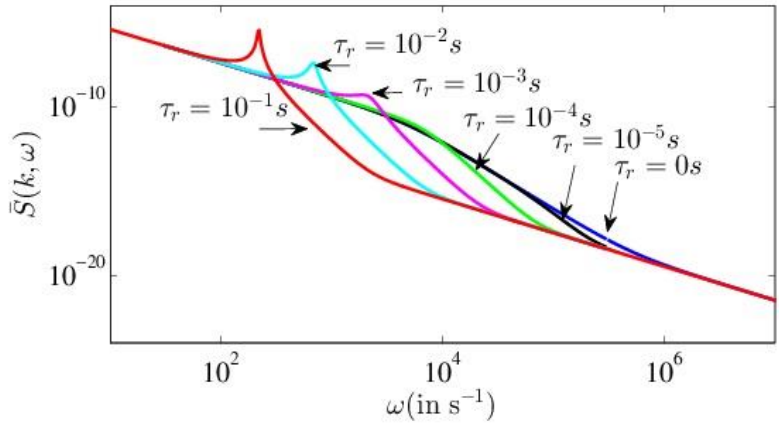

Figure 1. Contribution of the normalized dynamic structure factor $\overline{S}(k,\omega)$ towards the Rayleigh spectrum as a function of the angular frequency $(\omega)$ for different relaxation times , $(\tau_r)$ for the case of isothermal free-diffusion. The case of the Newtonian fluid is also denoted in the same figure for comparison. (Time instant, $t = 1000\,s$, $k = 85200\ m^{-1}$).

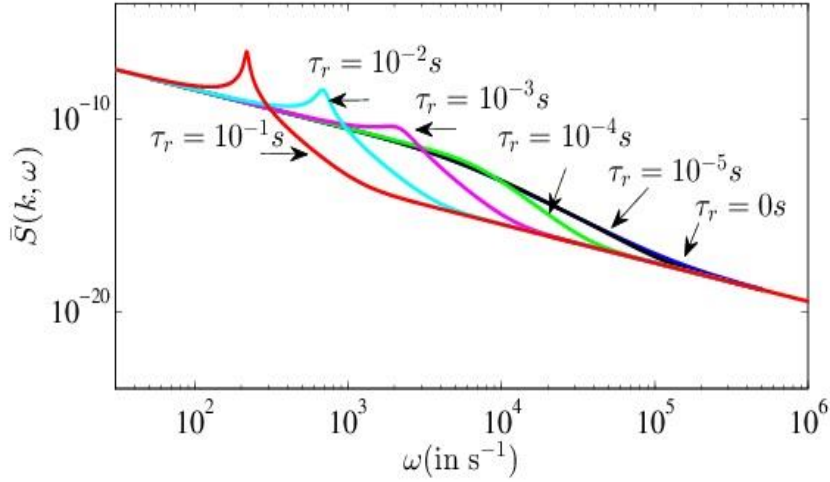

Figure 2. Contribution of the normalized dynamic structure factor $\bar{S}(k,\omega)$ towards the Rayleigh spectrum as a function of the angular frequency $(\omega)$ for different relaxation times $(\tau_r)$ for the case of thermal diffusion. The case of Newtonian fluid is also denoted in the same figure for comparison. (Time instant, $t = 1000\,s$, $k = 85200\ m^{-1}$).

Figures 1 and 2 depict the contribution of the normalized dynamic structure factor $\left(\bar{S}(k,\omega) = \int_0^a S_{k,\omega}\dfrac{(2\pi)^4}{2k_B T}dz\right)$ (please refer Eq. (26)) towards the Rayleigh spectrum as a function of the the angular frequency, for various relaxation times of the Maxwell fluid, for the cases of isothermal free diffusion and thermal diffusion respectively. From the figures, we see that for a given non-zero relaxation time, there is a peak followed by a gradual decreasing trend. As the relaxation time decreases (fluid tends to become more Newtonian in nature), the magnitude of the peak decreases, which is accompanied by a corresponding shift in the location of the peak towards larger frequencies. It should be noted that contribution towards the Rayleigh spectrum is from the transverse velocity fluctuations only. From the figures, it may be seen that $\bar{S}(k,\omega)$ is lower for a certain range of angular frequency for the Maxwell fluid as compared to the Newtonian fluid owing to the interplay of the Maxwell relaxation time with the angular frequency as dictated by the term $\dfrac{2k_B T}{(2\pi)^4(1+\omega^2\tau_r^2)}$. As $\tau_r \to 0$, not only is the peak absent, but the decay is also slower. The variation in the structure factor as a function of the angular frequency is closely related to the fluctuating kinetic energy contributions as we shall see later. Before that, we briefly look at the different regimes in the variation of the Structure factor. Initially, $\omega^{-2}$ behavior is observed for the variation of the structure factor as seen in figures 1 and 2, for both the cases of Maxwell and Newtonian characteristics. The concentration fluctuations ultimately decay with $\omega^{-2}$ behavior, which is observed in figures 1 and 2.

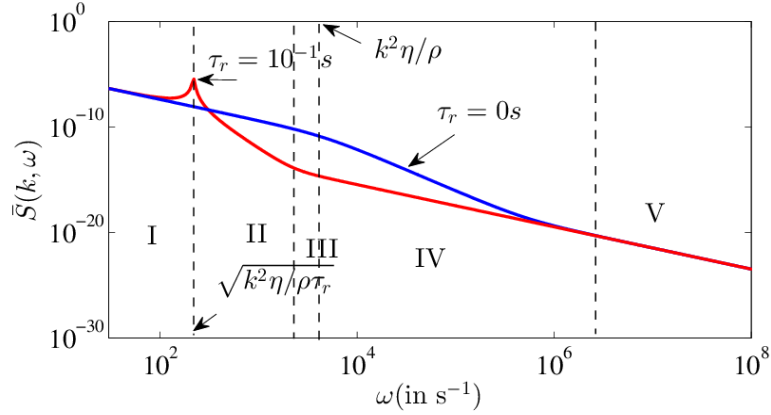

Figure 3. Various regimes present in the plot of normalized overall dynamic structure factor and $\omega$ for the case of Rayleigh spectrum (plotted for the case of free diffusion). (Time instant, $t$ =1000 $s$, $k$ =85200 $m^{-1}$).

Figure 3 depicts the variation of the normalized overall dynamic structure factor as a function of the angular frequency. In region I, as we move towards higher values of $\omega$, we observe that the intensity of concentration fluctuations increase. This can be attributed to the increasing intensity of velocity fluctuations. The underlying mechanism is that while there is continuous deformation (strain rate) in the system, some part of the energy is dissipated in the system, whereas some part of it is stored as elastic potential energy, for the case of Maxwell fluid. At some time scale of observation, some part of this potential energy is used up towards increasing the kinetic energy (related to the intensity of velocity fluctuations) and the rest gets dissipated in the system. The peaks are observed roughly at $\omega = \sqrt{\dfrac{\eta k^2}{\rho \tau_r}}$ (at the interface of region I and II). Evidently, the intensity cannot increase indefinitely, and at some time scale the natural intrinsic source of stochastic forcing, the random force given by Eq. (23), dominates. The velocity fluctuations cannot be sustained for very large values of $\omega$, as the random force responsible for the velocity fluctuation varies as $(\omega \tau_r)^{-2}$. Hence, the intensity of concentration fluctuations decay, as we move further in region II. Region III depicts that velocity fluctuations solely cannot sustain the concentration fluctuations and the random flux term given by Eq. (23)

prevails, which is responsible for the $\omega^{-2}$ behavior. The frequency at which the curve for the Maxwell fluid intersects the Newtonian curve while the intensity of decay in concentration fluctuations is given roughly by $\omega = \sqrt{\dfrac{2\eta k^2}{\rho \tau_r}}$ .

Another subtle observation can be made in the region between III and IV. The Newtonian curve deviates from its linear nature, and gradually transfers towards the region of lesser intensity of fluctuations, as have been shown in the entire region IV. The transition occurs roughly at $\omega = \eta k^2 / \rho$. The reason is that, as $\omega$ increases, the time scale of observation reduces till we reach a point where it coincides with the relaxation time for momentum diffusion. If the value of $\omega$ is further increased, momentum does not get enough time to diffuse and the concentration fluctuations are primarily dominated by the term given in Eq. (23).

From Eq. (20), the autocorrelation function of velocity fluctuations is related to the kinetic energy of fluctuations [35][37]. The auto-correlation of transverse component, namely. $\left\langle \delta \vec{u}_t . \delta \vec{u} *_t \right\rangle$ , yields:

$$\left[\frac{\eta k^2}{\rho^2(1+\omega^2 \tau_r^2)}\right] \left| \frac{\beta\left(\dfrac{g}{\left(i\omega + Dk^2\right)\left(i\omega + \eta X k^2\right) + \beta\left(\overline{g}.\nabla c\right)}\right) - 1}{\left(i\omega + \eta X k^2\right)}\right|^2$$

$$+ \frac{(\beta g k)^2 \dfrac{D}{\rho}\left(\dfrac{\partial c}{\partial \mu}\right)_{p,T}}{\left|\left(i\omega + Dk^2\right)\left(i\omega + \eta X k^2\right) + \beta\left(\overline{g}.\nabla c\right)\right|^2} \tag{37}$$

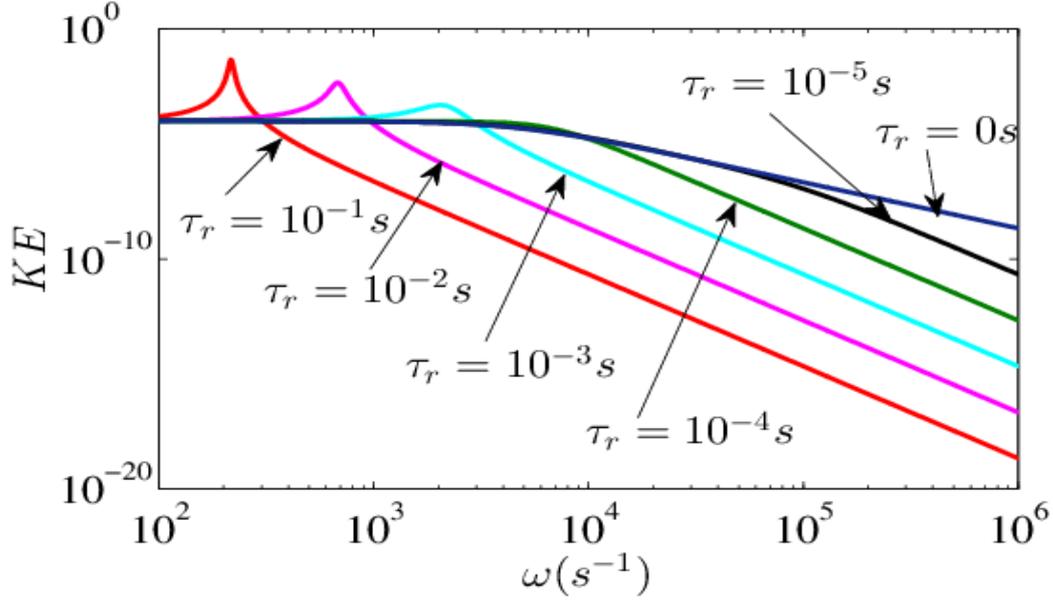

Figure 4. Plot of normalized kinetic energy due to the transverse velocity

fluctuations, $\text{KE} = \int\limits_{0}^{a} \left\langle \delta\vec{u}_t . \delta\vec{u}*_t \right\rangle \frac{(2\pi)^4}{(2k_B T)} dz$ versus $\omega$.

(Time instant, $t = 1000$ s, $k = 85200$ $m^{-1}$).

Figure 4 depicts the enhancement of kinetic energy, over a range of $\omega$. We observe that for lower values of $\omega$, the kinetic energy remains roughly constant for the case of Newtonian fluid, and then exhibits a decaying nature, where the decay is faster for the case of Maxwell fluid, owing to the $(\omega\tau_r)^{-2}$ nature of the variation of the random force responsible for velocity fluctuations as seen in Eq. (23). However, for the case of Newtonian fluid, the kinetic energy remains constant till $\omega = \eta k^2 / \rho$ (see figure 4), after which the kinetic energy decreases, which can be attributed to the fact that momentum does not get enough time to diffuse. We do observe peaks at $\omega = \sqrt{\dfrac{\eta k^2}{\rho\tau_r}}$, which strongly quantifies that velocity fluctuations dominates the nature of variation of concentration fluctuations. The situation is different for very small time instances.

*Short Time Dynamics.* Having studied the long time behavior of the structure factor for the Maxwell fluids, we proceed further to depict the evolution of the dynamic structure factor in the initial transients. Figures 5 and 6 depict the variation of normalized dynamic structure factor as a function of $\omega$ for the case of free diffusion.

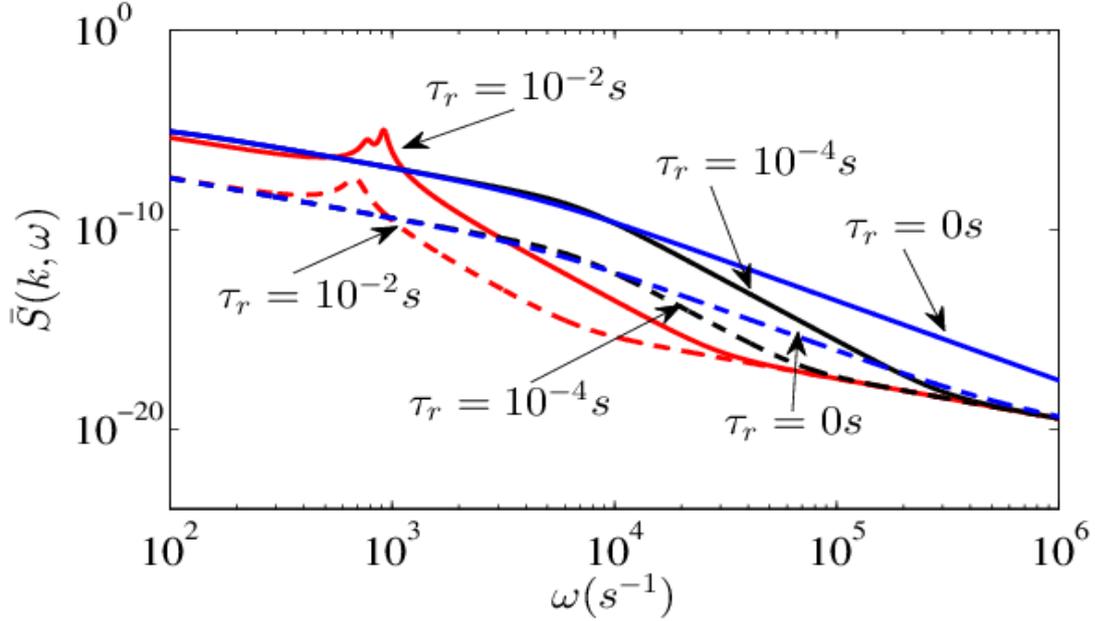

Figure 5. Contribution of the normalized dynamic structure factor $\bar{S}(k,\omega)$ towards the Rayleigh spectrum as a function of the angular frequency $\omega$ for different relaxation times $\tau_r$ for the case of free diffusion. The case of Newtonian fluid is also denoted in the same figure for comparison (solid lines: $t = 10^{-2}\,s$, dashed lines: $t = 10^{4}\,s$, $k = 85200\ m^{-1}$).

For figure 5, we observe the presence of time dependent peak(s), the occurrence of which is independent on rheology (though interestingly, only the intensity and the location of the peak is dependent upon $\tau_r$). An interesting observation is that the peak appears only for very small time instances and vanishes as time progresses. This may be attributed to the fact that initially there exists a very sharp concentration gradient at the interface (at the mid-height) between two fluids of different densities (shown in figures A1 and A3 in the appendix). The term $\beta(\vec{g}.\nabla c)$ becomes significant in the denominator of Eq. (26), which gives rise to a peak that is time-dependent in nature.

In figure 5, we observe that for that specific instant, the peak which we observed at $\omega = \sqrt{\dfrac{\eta k^2}{\rho \tau_r}}$ for large time instances is absent for Maxwell fluids. Two important parameters that have contextual relevance in this regard are [5]): $\tau_{diff} = 1/Dk^2$, which is the diffusion time constant, and $\tau_{grav} = \dfrac{\eta k^2}{\rho \beta(\vec{g}.\nabla c)}$. These parameters are present in the denominator of Eq. (26). Now, the concentration fluctuations are linked to the density fluctuations. We observe the involvement of buoyancy force, due to the influence of which, the concentration fluctuations involved will try to move the fluid towards the density matching layer. Meanwhile, the excess concentration is disposed off by the process of diffusion. For short time instances, the ratio $\tau_{diff} / \tau_{grav}$ is large. This implies that for very short time instances, diffusion is a slow process when compared to the buoyancy driven flow, and hence we observe the peaks in the intensity of concentration fluctuations. On the other hand, for large time instances, $\tau_{diff} / \tau_{grav}$ is small, and hence diffusion gets sufficient time to smear off the excess concentration, which leads to the disappearance of the time-dependent peaks for long time instances. Interestingly, the location of the peak is modified due to the presence of the factor $\rho X$, where $\rho X = \dfrac{(1 - i\omega\tau_r)}{(1 + \omega^2 \tau_r^2)}$ for non-zero values of $\tau_r$, indicating the influence of the fluid relaxation time.

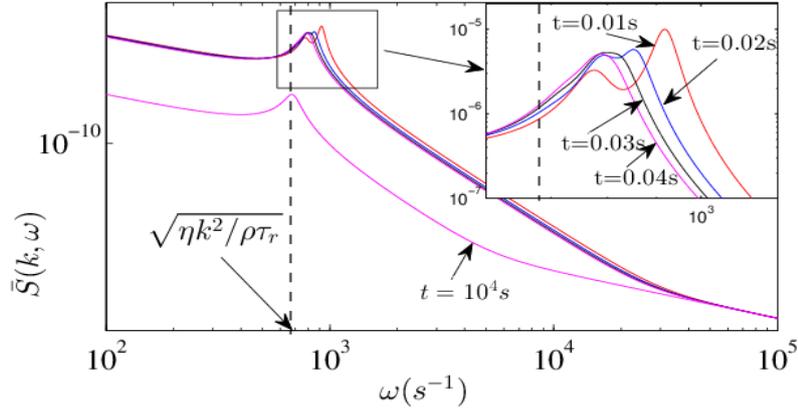

Figure 6. Contribution of the normalized dynamic structure factor $\bar{S}(k,\omega)$ towards the Rayleigh spectrum as a function of the angular frequency $\omega$ for the relaxation time $\tau_r = 10^{-2}$ $s$ for the case of free diffusion. The time-dependent nature of the peak is shown. ($k$ =852 $m^{-1}$).

Figure 6 depicts the variation of contribution of the normalized dynamic structure factor as a function of $\omega$, for different time instances ($t$ =0.01, 0.02, 0.03, 0.04 and $10^4$ $s$). It can be observed that at extremely small time instances, the time-dependent peak occurs for a higher frequency. As the time progresses, the peak starts shifting towards a lower frequency limit. The long time limit is characterized by the transverse velocity fluctuations and occurs at $\omega = \sqrt{\dfrac{\eta k^2}{\rho \tau_r}}$. This disappears for short-time instances. This can be attributed to the term $\beta(\vec{g}.\nabla c)$ [refer Eq. (26)], which becomes dominant for very small time instances as had already been discussed earlier.

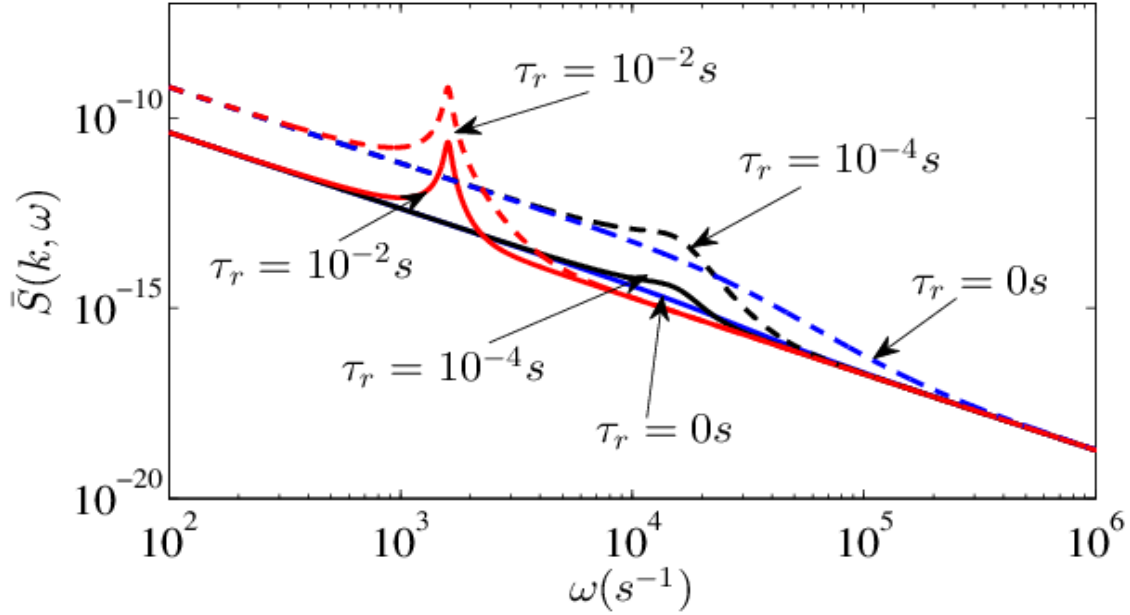

Figure 7. Contribution of the normalized dynamic structure factor $\bar{S}(k,\omega)$ towards the Rayleigh spectrum as a function of the angular frequency $(\omega)$ for different relaxation times $(\tau_r)$ for the case of thermal diffusion. The case of the Newtonian fluid is also denoted in the same figure for comparison. Time instants, (dashed lines) $=10^3 s$, (solid lines) $=10^{-2} s$. ($k = 2 \times 10^5\ m^{-1}$)

Figure 7 depicts the variation of normalized dynamic structure factor as a function of $\omega$, for the case of thermal diffusion. We do not observe the presence of any time-dependent peak, which was observed in the case of free diffusion, but the intensity of concentration fluctuations is found to be time-dependent, with the presence of the peaks due to the influence of rheology. As shown in figures A2 and A4 (please refer to the Appendix), we do not have the presence of any sharp concentration gradients at the mid-height (although we have sharp gradients at the boundaries for very small time instances), which limits the buoyancy driven flow (contributed by the term, $\beta\left(\vec{g}.\nabla c\right)$). Accordingly, we do not observe the presence of any time dependent peak for the case of thermal diffusion.

Having obtained interesting theoretical results, we must note that typical light scattering experiments use photon-correlation techniques in particular while investigating Rayleigh line. The minimum time delay is of the order of 0.1ms. To go to smaller time scales, require usage of techniques such as cross-correlation.

*Statics.* Typically, it has been shown [5] that at large wave vectors, the static structure factor displays the $k^{-4}$ behavior, while at lower values it assumes a constant value. For completeness, we present the effect of rheology on static structure factor. The static structure factor as opposed to the dynamic structure factor can be defined as [5, 12], $\bar{S}(\mathrm{k}) = \int\limits_{-\infty}^{\infty} \bar{S}(\mathrm{k}, \omega) d\omega$. Our analysis shows that rheology does not have drastic influence as opposed to the case of dynamic structure factor, but after the saturation regime, which is roughly the same for all the cases of $\tau_r$, the non-Newtonian curves decay slowly as compared to the Newtonian curve, although for $\tau_r > 10^{-2} s$, the curves roughly become identical to the Newtonian curve. While for the case of thermal diffusion, we did not observe any significant difference.

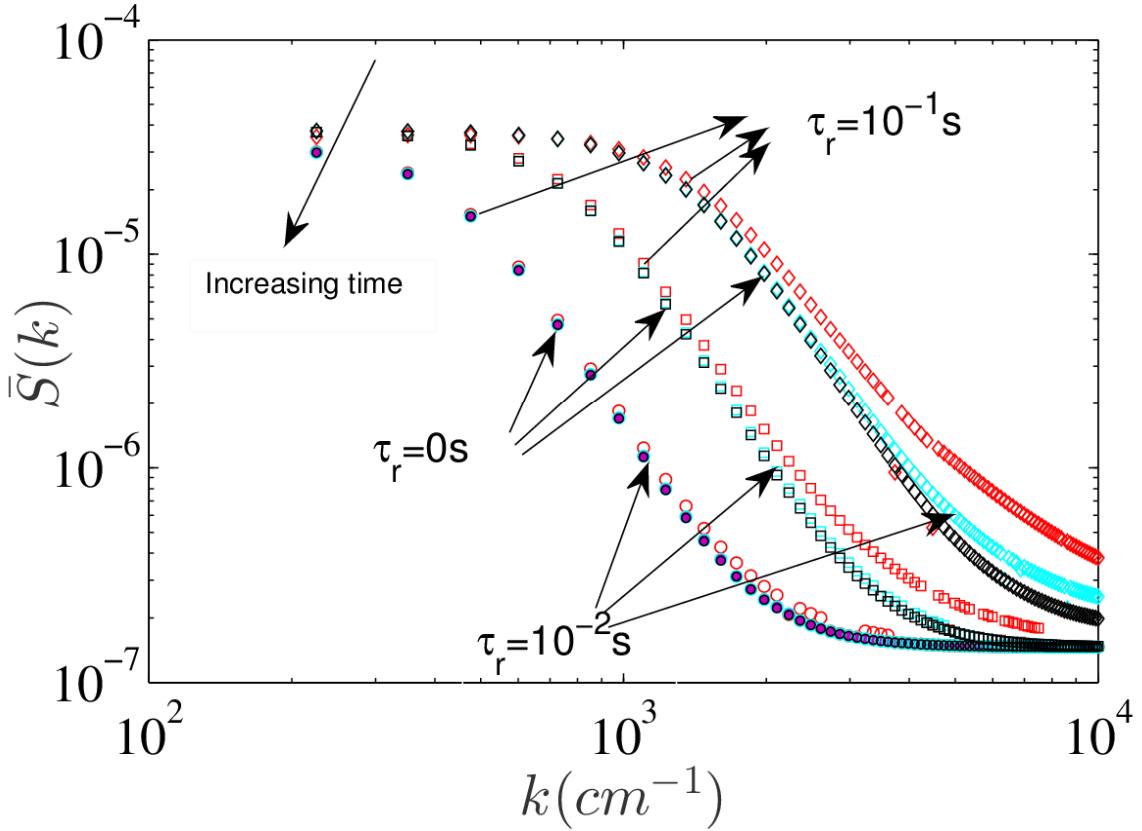

Figure 8. Static structure factor $\bar{S}(k)$ for different relaxation times $(\tau_r)$ for the case of free diffusion. The case of the Newtonian fluid is also denoted in the same figure for comparison. Time instant(s) =1, $10^2$ and $10^4 s$.

Our analysis proves the fact that non-equilibrium fluctuations are indeed affected by viscoelastic effect through its influence on both the static and dynamic structure factor. For mathematical clarity, we present a calculation as follows: If we look at equation (6) and replace g (acceleration due to gravity) with 0 (that is the system is not under the influence of gravity), we get,

$$< \delta c_{k,\omega} \delta c_{k,\omega}^* >_R = \frac{2k_B T}{(2\pi)^4} k^2 \left[ \frac{D}{\rho} \left( \frac{\partial c}{\partial \mu} \right)_{p,T} \frac{1}{\left( \omega^2 + D^2 k^4 \right)} + \frac{\frac{(\eta \mid \nabla c \mid^2)}{(1+\omega^2 \tau_r^2) \rho^2}}{\mid (i\omega + Dk^2)(i\omega + \eta X k^2) \mid^2} \right] \quad [38]$$

This can be expressed as the sum of Lorentzians as follows:

$$< \delta c_{k,\omega} \delta c_{k,\omega}^* >_R = \frac{2k_B T}{(2\pi)^4} k^2 \left[ \frac{D}{\rho} \left( \frac{\partial c}{\partial \mu} \right)_{p,T} \frac{1}{\left( \omega^2 + D^2 k^4 \right)} + \frac{\frac{(\eta \tau_r^2 \mid \nabla c \mid^2)}{\rho^2}}{\left( \omega^2 + D^2 k^4 \right) \left( \omega^2 \tau_r^2 + a^2 \right) \left( \omega^2 \tau_r^2 + b^2 \right)} \right] [39]$$

where, $a = \frac{1}{2} \left( 1 - \sqrt{1 - \frac{4\eta \tau_r k^2}{\rho}} \right)$ and $b = 1 - a$. Now the decomposition into three Lorentzians is valid as long as a, b is real. This gives us, $\frac{4\eta \tau_r k^2}{\rho} < 1$. If this condition is not satisfied, then the modes associated with $\tau_r$ become propagative. Then equation (38) can be written as,

$$< \delta c_{k,\omega} \delta c_{k,\omega}^* >_R = \frac{2k_B T}{(2\pi)^4} k^2 \left[ \frac{D}{\rho} \left( \frac{\partial c}{\partial \mu} \right)_{p,T} \frac{1}{\left( \omega^2 + D^2 k^4 \right)} + \frac{\frac{(\eta \tau_r^2 \mid \nabla c \mid^2)}{\rho^2}}{\left( \omega^2 + D^2 k^4 \right) \left( \left( \omega - a_p \right)^2 + b_p^2 \right) \left( \left( \omega + a_p \right)^2 + b_p^2 \right)} \right]$$
$$[40]$$

where, $a_p$ gives the (non-zero) location of the peak. We observe that the structure factor can be decomposed in two parts, an equilibrium and a non-equilibrium part. Equation (40) indeed concludes that at equilibrium, that is, when $\nabla c = 0$, we do not get viscoelastic effects in the Rayleigh line. Now the situation is complicated for the case when gravity is present and no simple analytical expression is possible. But, in our analysis, we have already investigated that peaks, (at large times) do appear at $\omega = \sqrt{\frac{\eta k^2}{\rho \tau_r}}$, whereas, at short-times for the case of free diffusion depends on $\div t \phi$ (time), which essentially makes us conclude that the location of the peak(s) (please refer figure 6) at all times is a function of $\sqrt{\dfrac{\eta k^2}{\rho \tau_r}}$ and $-t\phi$ (time) (which can be manifested only through the presence of $\nabla c$). Intriguingly, for the case of $g \neq 0$ and large time instances, the criteria for appearance of propagative mode remains roughly the same as for the case of $g = 0$. [This can be easily verified if one takes the final normalized structure factor (as shown in figures of dynamic structure factor) and normalizes it again with respect to its individual peak value. These curves for large time instances do become time-independent (implying that $b_p$ from equation (40) also remains roughly independent of $\nabla c$) which makes us conclude that indeed the structure factor behaves as if the denominator of equation (26) becomes independent of $\nabla c$]. However, for the case of $g \neq 0$ and shorter time instances, the criteria for appearance of propagative mode becomes complicated and no simple analytical expression is possible.

## IV. CONCLUSIONS

Our study indicates that rheology indeed plays a dominant role in the non-equilibrium behavior of concentration fluctuations. Rheology affects transverse velocity fluctuations, quantified by the velocity auto-correlation function (dynamic structure factor) which in turn affects the nature of variation of the concentration fluctuations for large time instances. At small times, rheology has its influence on the part contributing towards the Rayleigh spectrum in the form of the transverse component of velocity fluctuations. The spectra for short time instances are dominated by the presence of buoyancy force for the case of free diffusion, but have no role to play for the case of thermal diffusion for the specific arrangement of the fluid mixture considered in our case. The analysis for static structure factor reveals a weak dependence on rheology especially at large time instances.

## ACKNOWLEDGEMENT

The authors gratefully acknowledge the anonymous referee for the insightful discussion on the zero-buoyancy case and for the physical insight for obtaining the nonequilibrium components of the spectrum in a more illuminating manner.

## APPENDIX: TIME EVOLUTION OF CONCENTRATION PROFILES

Spatio-temporal evolution of the concentration and the concentration gradient

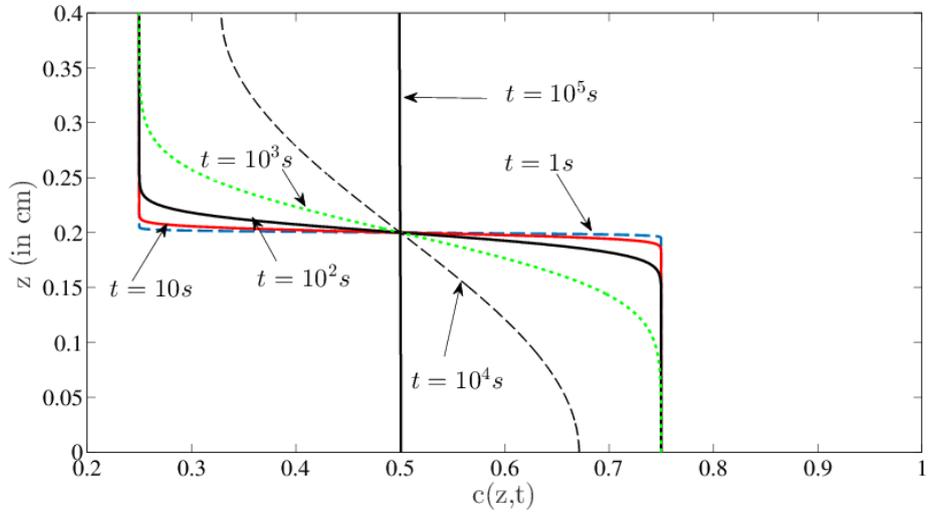

Figure A1. Evolution of the concentration profile versus the vertical height for the case of free diffusion.

Figure A1 depicts the variation of the concentration as a function of time for the case of free diffusion at $t = 10^n\ s$; $n = 0, 1 \ldots 5$. The concentration profiles are depicted for the expression (3). The other parameters appear in the opening paragraph of the results and discussion section.

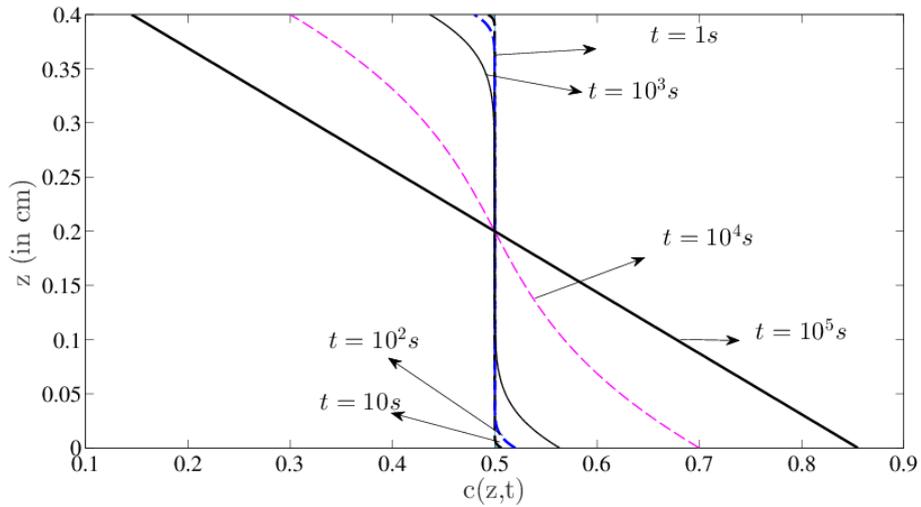

Figure A2. Evolution of the concentration profile versus the vertical height for the case of thermal diffusion.

Figure A2 depicts the variation of the concentration as a function of time for the case of thermal diffusion at $t = 10^n\ s$; $n = 0, 1 \ldots 5$. The concentration profiles are depicted for the expression (35).

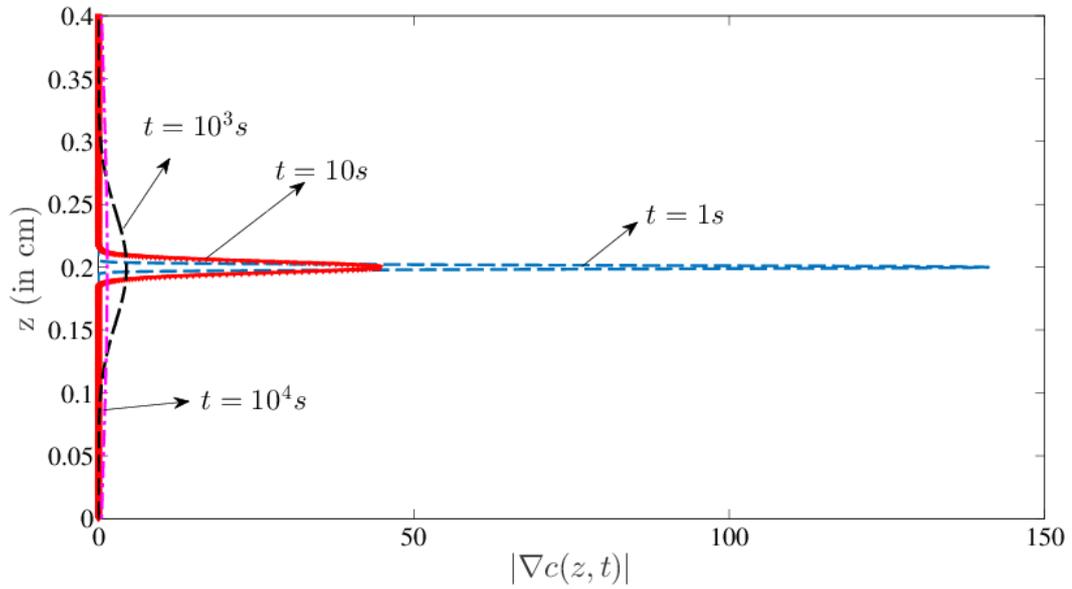

Figure A3. Evolution of the gradient of the concentration profile versus the vertical height for the case of free diffusion.

Figure A3 depicts the variation of the concentration gradient as a function of time for the case of free diffusion at t = $10^n$ s; n = 0, 1 í   4. The concentration gradient profiles are depicted for the expression (30), corresponding gradient to that expression. The plots roughly match with expression (31) for the given time instances.

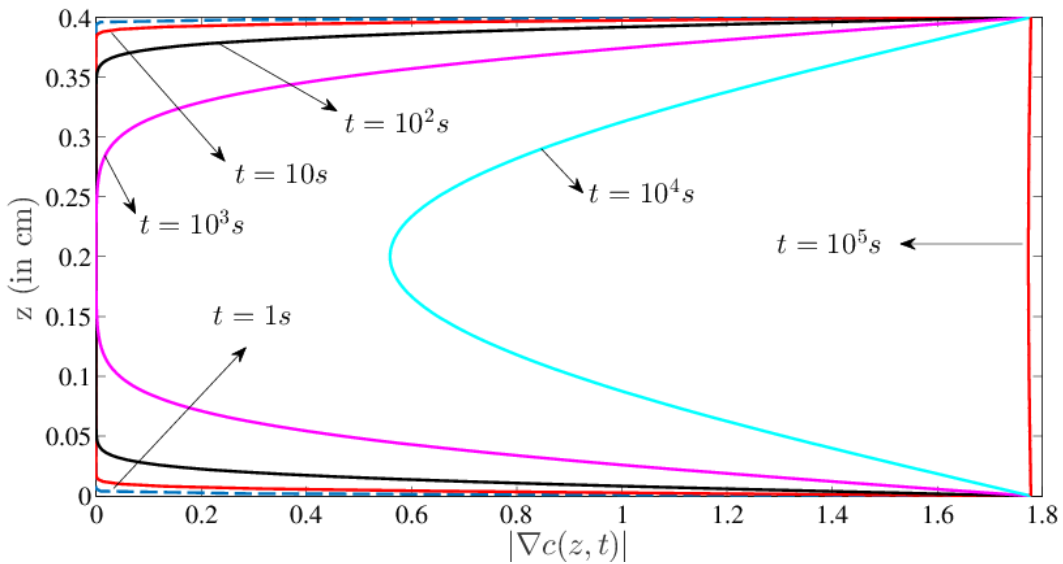

Figure A4. Evolution of the gradient of the concentration profile versus the vertical height for the case of thermal diffusion.

Figure A4 depicts the variation of the concentration gradient as a function of time for the case of thermal diffusion at $t = 10^n\ s$; $n = 0, 1.\ \text{í}\quad 5$. The concentration profiles are depicted for the expression (36) which follows from (35).